\newcount\Comments  
\documentclass[fleqn,usenatbib]{mnras}

\usepackage{newtxtext,newtxmath}

\usepackage[T1]{fontenc}

\DeclareRobustCommand{\VAN}[3]{#2}
\let\VANthebibliography\thebibliography
\def\thebibliography{\DeclareRobustCommand{\VAN}[3]{##3}\VANthebibliography}

\usepackage{graphicx}	
\usepackage{amsmath}	
\usepackage{dcolumn}
\newcolumntype{.}{D{.}{.}{-1}}
\newcolumntype{u}{D{,}{\;\pm\;}{-1}}

\usepackage{color}
\definecolor{darkgreen}{rgb}{0,0.5,0}
\definecolor{purple}{rgb}{1,0,1}
\newcommand{\kibitz}[2]{\ifnum\Comments=1\textcolor{#1}{#2}\fi}

\newcommand{\changes}[1]{{#1}}

\usepackage[switch]{lineno}

\newcommand{\per}{$^{-1}$}
\newcommand{\degree}{$^{\circ}$}

\title[MERGHERS pilot study]{{\color{black}MERGHERS Pilot: MeerKAT discovery of diffuse emission in nine massive Sunyaev-Zel'dovich-selected galaxy clusters from ACT}}

\author[K. Knowles et al.]{\Large K.~Knowles,$^{1,2}$\thanks{E-mail: kendaknowles.astro@gmail.com (KK)}
D.\,S.~Pillay,$^{1,2}$
S.~Amodeo,$^{3}$
A.\,J.~Baker,$^{4}$
K.~Basu,$^{5}$
D.~Crichton,$^{6}$
F.~de~Gasperin,$^{7,8}$
\newauthor\Large M.~Devlin,$^{9}$
C.~Ferrari,$^{10}$
M.~Hilton,$^{1,2}$
K.\,M.~Huffenberger,$^{11}$
J.\,P.~Hughes,$^{4}$
B.\,J.~Koopman,$^{12}$
K.~Moodley,$^{1,2}$
\newauthor\Large T.~Mroczkowski,$^{13}$
S.~Naess,$^{14}$
F.~Nati,$^{15}$
L.\,B.~Newburgh,$^{12}$
N.~Oozeer,$^{16,17}$
L.~Page,$^{18}$
B.~Partridge,$^{19}$
\newauthor\Large C.~Pfrommer,$^{20}$
M.~Salatino,$^{21,22}$
A.~Schillaci,$^{23}$
C.~Sif\'{o}n,$^{24}$
O.~Smirnov,$^{25,16}$
S.\,P.~Sikhosana,$^{1,2}$
\newauthor\Large E.\,J.~Wollack,$^{26}$
Z.~Xu$^{9,27}$
\\
$^{1}$Astrophysics Research Centre, University of KwaZulu-Natal, Durban 4041, South Africa\\
$^{2}$School of Mathematics, Statistics \& Computer Science, University of KwaZulu-Natal, Westville Campus, Durban 4041, South Africa\\
$^{3}$Department of Astronomy, Cornell University, Ithaca, NY 14853, USA\\
$^{4}$Department of Physics and Astronomy, Rutgers, the State University of New Jersey, 136 Frelinghuysen Rd., Piscataway, NJ 08854-8019, USA\\
$^{5}$Argelander Institute for Astronomy, University of Bonn, Auf dem Huegel 71, 53121 Bonn, Germany\\
$^{6}$Institute for Particle Physics and Astrophysics, Eidgen\"ossische Technische Hochschule Z\"urich, Wolfgang-Pauli-Str. 27, 8093 Z\"urich, Switzerland\\
$^{7}$Hamburger Sternwarte, University of Hamburg, Gojenbergsweg 112, D-21029, Hamburg, Germany\\
$^{8}$Istituto di Radioastronomia, via P. Gobetti 101, 40129, Bologna, Italy\\
$^{9}$Department of Physics and Astronomy, University of Pennsylvania, 209 South 33rd Street, Philadelphia, PA 19104, USA\\
$^{10}$Laboratoire Lagrange, Université Cote d'Azur, Observatoire de la Cote d'Azur, CNRS, F-06300, Nice, France\\
$^{11}$Department of Physics, Florida State University, Tallahassee, FL 32306, USA\\
$^{12}$Department of Physics, Yale University, New Haven, CT 06520, USA\\
$^{13}$European Southern Observatory, Karl-Schwarzschild-Str. 2, D-85748 Garching b. München, Germany\\
$^{14}$Center for Computational Astrophysics, Flatiron Institute, New York, NY 10010, USA\\
$^{15}$Department of Physics, University of Milano-Bicocca, Piazza della Scienza 3, 20126 Milano (MI), Italy\\
$^{16}$South African Radio Astronomy Observatory, 2 Fir Street, Black River Park, Observatory, Cape Town 7925, South Africa\\
$^{17}$African Institute for Mathematical Sciences, 6 Melrose Road, Muizenberg 7945, South Africa\\
$^{18}$Joseph Henry Laboratories of Physics, Jadwin Hall, Princeton University, Princeton, NJ 08544, USA\\
$^{19}$Department of Physics and Astronomy, Haverford College,Haverford, PA 19041, USA\\
$^{20}$Leibniz-Institute for Astrophysics Potsdam (AIP), An der Sternwarte 16, 14482 Potsdam, Germany\\
$^{21}$Kavli Institute for Particle Astrophysics and Cosmology (KIPAC), Stanford, CA 94305, USA\\
$^{22}$Stanford University, Stanford, CA 94305, USA\\
$^{23}$Department of Physics, California Institute of Technology, Pasadena, CA 91125, USA\\
$^{24}$Instituto de F\'isica, Pontificia Universidad Cat\'olica de Valpara\'iso, Casilla 4059, Valpara\'iso, Chile\\
$^{25}$Department of Physics and Electronics, Rhodes University, PO Box 94, Makhanda, 6140, South Africa\\
$^{26}$NASA/Goddard Space Flight Center, 8800 Greenbelt Rd, Greenbelt, MD 20771, USA\\
$^{27}$MIT Kavli Institute, Massachusetts Institute of Technology, Cambridge, MA, USA
}

\date{Accepted XXX. Received YYY; in original form ZZZ}

\pubyear{2020}

\begin{document}
\label{firstpage}
\pagerange{\pageref{firstpage}--\pageref{lastpage}}
\maketitle

\begin{abstract}
The MeerKAT Exploration of Relics, Giant Halos, and Extragalactic Radio Sources (MERGHERS) survey is a planned project to study a large statistical sample of galaxy clusters with the MeerKAT observatory. Here we present the results of a 16--hour pilot project, observed in response to the 2019 MeerKAT Shared Risk proposal call, to test the feasibility of using MeerKAT for a large cluster study using short (0.2--2.1\,hour) integration times. The pilot focuses on 1.28\,GHz observations of 13 massive, low-to-intermediate redshift ($0.22 < z < 0.65$) clusters from the Sunyaev-Zel'dovich-selected Atacama Cosmology Telescope (ACT) DR5 catalogue that show multiwavelength indications of dynamical disturbance. With a 70 per cent detection rate (9/13 clusters), this pilot study validates our proposed MERGHERS observing strategy and provides twelve detections of diffuse emission, eleven of them new, indicating the strength of MeerKAT for such types of studies. The detections (signal-to-noise ratio $\gtrsim6$) are summarised as follows: two systems host both relic(s) and a giant radio halo, five systems host radio halos, and two have candidate radio halos. \changes{Power values, $k$-corrected to 1.4\,GHz assuming a fiducial spectral index of $\alpha = -1.3 \pm 0.4$, are consistent with known radio halo and relic scaling relations.}   
\end{abstract}

\begin{keywords}
galaxies: clusters: general -- radio: continuum: general
\end{keywords}



\section{Introduction}


In the last two decades, observations of diffuse, steep-spectrum, cluster-scale radio synchrotron emission have been used to study the physical link between the thermal and non-thermal components of the intracluster medium (ICM), constraining theories of cosmic-ray transport and magnetic field evolution within the ICM \citep[see reviews by][]{BrunettiJones2014,vanWeeren2019}. 
A strong link has been found to the dynamics of the host cluster, with the largest diffuse emission classes such as radio halos and relics found in massive, dynamically disturbed systems -- radio halos are thought to originate from merger-driven turbulence within the ICM, and relics have been related to the presence of cluster merger shocks or revived radio galaxy plasma \citep[see e.g.,][and references therein]{vanWeeren2019}. 
The radio halo and relic luminosities correlate with the host cluster's mass and thermal properties \citep[][]{Cassano2013,FdG2014}. However, cluster selection has been found to affect these relations: samples selected via their Sunyaev-Zel'dovich \citep[SZ;][]{SZ1972} signal show a higher diffuse emission detection rate than X-ray-selected samples \citep{Basu2012,SB2014,Zandanel2014,Bonafede2015,Cuciti2015}, which may be due to an X-ray selection bias towards relaxed, cool-core clusters \citep{AS2017}, or the different timescales of SZ versus X-ray signal boosting during mergers \citep{Randall2002boostsX,Poole2007,Wik2008boostsSZ}.

All statistical, uniformly-selected cluster samples used in radio halo studies have to date been restricted to high masses ($M_{500}\,\gtrsim\,6\,\times\,10^{14}\,{\rm M}_\odot$) and lower redshift ranges \citep[$\,0.1\,\lesssim\,z\,<\,0.4$; e.g.,][]{Venturi2007,Cuciti2015}, with current turbulent re-acceleration formation theories predicting a sharp drop in the occurrence of radio halos at higher redshift, where cluster magnetic fields are expected to be weaker \citep[][]{Cassano2006, BrunettiJones2014}. Diffuse emission has also been detected in some low-mass \citep[e.g.,][]{Knowles2016,Bernardi2016,Kale2017,HL2018A2146,Bruggen2018lowmass,Hoang2020} and high-redshift \citep[\color{black}$z > 0.5;$][]{Bonafede2009,Lindner2014, Pandey2013,Riseley2017,Knowles2019,Giovannini2020} systems. Most of these detections result from single-target programmes or small samples. A study of a large statistical sample that covers these extended mass and redshift ranges has not been carried out. Most recently, LOFAR imaging of a declination-{\color{black}selected} sample of nineteen $z\,>\,0.6$, SZ-selected clusters  
revealed high-redshift diffuse emission with similar radio powers to those in lower redshift systems \citep[][]{diGenarro2020}. This is an indication that our theories of the evolution of cosmic magnetic fields, and therefore of the production of diffuse emission, require further study. 

The next step in advancing our understanding of diffuse emission processes and imposing stronger constraints on formation models is to perform statistical studies on cluster samples that expand discovery space. Due to their much improved sensitivity, such a study has become possible using the new generation of Square Kilometre Array (SKA) precursor telescopes such as MeerKAT, LOFAR \citep{LOFAR}, and ASKAP \citep{ASKAP}. MeerKAT is currently the most sensitive instrument of its kind in the Southern hemisphere. This 64-element interferometer located in the South African Karoo region, described in \cite{Jonas2016}, \cite{Camilo2018}, and \cite{DEEP2}, operates in the S (1750--3500\,MHz), L (900--1670\,MHz), and UHF (580--1015\,MHz) bands. Its configuration makes it particularly well-suited to studies of diffuse cluster emission: its dense core, with $\sim$ 75 per cent of the antennas lying within a 1\,km radius, provides superior brightness sensitivity to extended structures, while its outer ring antennas provide a maximum baseline of 8\,km and therefore sufficient resolution to disentangle compact sources in all but the highest redshift targets.

Due to their redshift-independent selection functions, large area SZ surveys are a more efficient way to select massive clusters at any redshift compared to X-ray surveys with telescopes like \textit{ROSAT} \citep{ROSAT} or \textit{eROSITA} \citep{EROSITA}, and are therefore better suited to an expanded diffuse emission study. Large SZ cluster catalogues have been compiled by the \textit{Planck} satellite \citep{Planck2014, Planck2016}, and the ground-based Atacama Cosmology Telescope \citep[ACT;][]{Hasselfield2013, Hilton2018} and South Pole Telescope \citep{Bleem2015, Bleem2020}. Unlike the ground based telescopes which have higher resolution, \textit{Planck} misses clusters at low mass and at higher redshift due to beam dilution. ACT is undertaking its Advanced ACT \citep[AdvACT;][]{Henderson2016} survey of $\sim$\,18,000\,deg$^2$ of the Southern and Equatorial sky \citep{Naess2020}, with the first release of the cluster catalogue (ACT DR5\footnote{The ACT DR5 catalogue uses the $\sim$\,13,000\,deg$^2$ away from the Galactic plane to search for clusters.}) containing more than 4000 optically-confirmed clusters \citep[][]{2021ApJS..253....3H}.

The MERGHERS \citep[MeerKAT Exploration of Relics, Giant Halos, and Extragalactic Radio Sources;][]{MERGHERS} survey is a planned large-scale radio follow-up of ACT cluster targets that will be blind to the cluster dynamical state and use short integration times on MeerKAT. We present here the results of a pilot study of 13 ACT DR5 clusters, to test the feasibility of using MeerKAT for a large-scale cluster programme such as MERGHERS.

The paper is organised as follows. In Section \ref{sec:sampleobs} we introduce the sample and discuss the observations. The data processing methodology and imaging are described in Section \ref{sec:processing}. Results are presented in Section \ref{sec:results}, with concluding remarks in Section \ref{sec:conclusion}. In this paper we adopt a $\Lambda$CDM flat cosmology with $H_0 = 70$ km s{\per} Mpc{\per}, $\Omega_{\rm m} = 0.3$, and $\Omega_{\rm \Lambda} = 0.7$. 


\section{Cluster sample and observing}
\label{sec:sampleobs}

\begin{table*}
	\centering
	\caption{Cluster sample with observational and imaging details. The clusters are listed by epoch of observation, with the number of antennas used in {\color{black}each epoch's} observation. Cols: (1) ACT DR5 cluster name; (2) J2000 Right Ascension of the SZ peak; (3) J2000 Declination of the SZ peak; (4) Cluster redshift; (5) ACT SZ weak-lensing-calibrated mass; (6) Total MeerKAT time on target; (7) Percentage of MeerKAT data flagged during processing, which includes the known frequency ranges affected by satellites; (8) Full resolution central rms noise; (9) Full resolution synthesised beam: major axis, minor axis, and position angle; (10-11) LS map rms noise and resolution, respectively -- see Section \ref{sec:results} for details. 
	}
	\label{tab:sampleobs}
	\begin{tabular}{lrrc.cc.ccc} 
		\hline
		\multicolumn{1}{c}{(1)} & \multicolumn{1}{c}{(2)} & \multicolumn{1}{c}{(3)} & (4) & (5) & (6) & (7) & (8) & (9) & (10) & (11) \\
		Name & \multicolumn{1}{c}{RA J2000} & \multicolumn{1}{c}{Dec. J2000} & z & \multicolumn{1}{c}{$M_{500}$} & $t_{\rm src}$ & Flagged & \multicolumn{1}{c}{$\sigma_{\rm FR}$} & \multicolumn{1}{c}{$\theta_{\rm synth,FR}$} & \multicolumn{1}{c}{$\sigma_{\rm LS}$} & $\theta_{\rm synth,LS}$ \\
		(ACT-CL...) & \multicolumn{1}{c}{(deg)} & \multicolumn{1}{c}{(deg)} & & \multicolumn{1}{r}{($10^{14} M_\odot$)} & (min) & (\%) & \multicolumn{1}{r}{($\mu$Jy beam\per)} & (\arcsec,\,\arcsec,\,\degree) & \multicolumn{1}{r}{($\mu$Jy beam\per)} & (\arcsec) \\
		\hline 
		\multicolumn{3}{l}{\textit{Epoch A - 61 antennas}}\\
		J0013.3$-$4906           &  3.32748 & $-$49.11263 & 0.407 &  6.8 & 108 & 39.6 
		&  7.6 & 6.9,\,5.9,\,152.0 & 24 & 21 \\ 
		J0019.6$+$0336$^\dagger$ &  4.91085 &   3.60879 & 0.266 & 10.2 &  24 & 44.2 
		& 18.1 & 7.9,\,7.2,\,161.7 & 52 & 21 \\ 
		J0034.4$+$0225$^\dagger$ &  8.61022 &   2.42259 & 0.388 & 8.1 &  24 & 39.0 
		& 23.1 & 8.2,\,6.7,\,147.8 & 77 & 26\\ 
		J0040.8$-$4407$^\dagger$ & 10.20664 & $-$44.13242 & 0.350 & 10.3 &  24 & 39.7 
		& 31.8 & 7.2,\,6.1,\,157.7 & 260$^{a}$ & 23 \\ 
		J0046.4$-$3912           & 11.60191 & $-$39.20152 & 0.592 &  7.9 &  48 & 39.4 
		& 10.2 & 6.9,\,5.9,\,152.4 & 30 & 21\\ 
		
		\multicolumn{3}{l}{\textit{Epoch B - 58 antennas}}\\
		J0106.1$-$0618           & 16.54119 &  $-$6.31591 & 0.641 &  4.6 & 108 & 49.3 
		&  7.5 & 7.4,\,5.1,\,165.4 & 25 & 23 \\ 
		J0159.0$-$3413           & 29.75418 & $-$34.22213 & 0.413 &  9.1 &  24 & 48.6 
		& 12.3 & 6.7,\,5.6,\,163.6 & 35 & 21\\ 
		J0240.0$+$0115$^\dagger$ & 40.01278 &   1.26642 & 0.603 &  5.0 & 132 & 50.6 
		&  8.1 & 7.7,\,5.1,\,164.9 & 27 & 23 \\ 
		
		\multicolumn{3}{l}{\textit{Epoch C - 58 antennas}}\\
		J0245.5$-$5302           & 41.37543 & $-$53.03602 & 0.298 & 10.7 &  24 & 61.3 
		& 14.7 & 7.7,\,5.1,\,\,16.3 & 39 & 23\\ 
		J0248.1$-$0216           & 42.04673 &  $-$2.27442 & 0.238 & 9.9 &  24 & 47.9 
		& 13.4 & 7.7,\,5.4,\,165.9 & 41 & 23 \\ 
		J0248.2$+$0238$^\dagger$ & 42.05431 &   2.63611 & 0.554 &  6.8 &  72 & 52.0 
		& 14.0 & 8.3,\,5.2,\,160.1 & 46 & 21 \\ 
		J0528.8$-$3927           & 82.21643 & $-$39.46265 & 0.284 &  9.0 &  39 & 47.4 
		& 10.4 & 6.9,\,5.0,\,157.5 & 37 & 21 \\ 
		J0638.7$-$5358           & 99.69631 & $-$53.97338 & 0.226 & 12.5 &  24 & 50.0 
		& 15.5 & 7.8,\,5.4,\,146.5 & 48 & 23 \\ 
		\hline
	\end{tabular}
	
	Notes:\,\,$^\dagger$ Cluster field required direction-dependent corrections. $^a$ Image has highly variable noise due to residual contamination by the bright source artefacts.
\end{table*}

\subsection{Sample selection}
\label{subsec:sample}
For our cluster selection, we used an early version of the ACT DR5 cluster catalogue based on the ACT data through 2016, and selected candidate massive mergers to ensure a high scientific return from the pilot project. The full MERGHERS sample will not be subject to this ``candidate merger'' constraint, as it will be homogeneously mass-selected. 

The selection criteria for the pilot sample were as follows. To ensure that our targets were robust cluster detections, we restricted the preliminary ACT DR5 sample to optically confirmed clusters\footnote{We refer the reader to \citet{2021ApJS..253....3H} for the optical confirmation procedure.} with an SZ signal-to-noise ratio (SNR) greater than 10, with a further Right Ascension (RA) range cut of 23h to 07h. The RA restriction was based on an expected MeerKAT observing schedule ranging from July through November, and on our preference for night-time observing. The SNR cut ensures we selected high-mass systems, and we further constrained the selection to clusters with $z\,\lesssim\,0.6$ that lie within the coverage of the Dark Energy Survey \citep[DES;][]{DES}. These cuts reduced the sample pool to 29 clusters. To select potential mergers, we used a qualitative optical indication of disturbed morphology by calculating the positional offset between the SZ peak and the brightest cluster galaxy (BCG); the larger the offset, the more likely it is that the cluster is disturbed \citep{Sehgal2013}. We visually inspected the DES imaging to confirm the presence of multiple BCGs for clusters with significant or intermediate offsets. 

The required integration time per target was determined through mock MeerKAT observations and a goal MeerKAT detection SNR of 10. The mock observations conservatively assumed 400\,MHz of useable bandwidth due to the known satellite-affected frequency ranges. The simulations assumed a halo detection with a surface brightness, given the cluster redshift and SZ-derived mass\footnote{We use the weak-lensing-calibrated ACT masses.}, determined from observed scaling relations \citep{Cassano2013}. Our final sample was selected so as to maximise the number of targets observed, while remaining, after observation overheads, within the 16 hour time restriction of the 2019 MeerKAT Shared Risk proposal call. The list of 13 ACT DR5 clusters observed in this pilot study, along with the on-target times, is provided in Table \ref{tab:sampleobs}. 


\subsection{Observations}
\label{subsec:strategy}

All data for this project were observed in MeerKAT's L-band receiver configuration, which has a native bandwidth of 856\,MHz and a central frequency of 1.28\,GHz. We observed in full polarization using the 4096 channel mode and 8\,s dump time. To make full use of the power of MeerKAT, all observations used at least 58 of the 64 antennas; we also required that at least seven of the nine outer ring antennas be included. This criterion ensured that we would retain sufficient resolution in the final images to disentangle compact sources from any observed diffuse emission: for the highest redshift cluster in our sample, $z\,=\,0.640$, a physical scale of 200\,kpc (a conservative lower limit for merger-related diffuse cluster emission) covers $\sim$\,29\,arcsec on the sky. 

At L-band, a noise level of 10\,$\mu$Jy\,beam{\per}, using Briggs \citep[][]{BRIGGS} robust 0 weighting for a $\sim$\,10\,arcsec beam, can be reached with less than an hour of MeerKAT time\footnote{See the online MeerKAT sensitivity calculator at \url{https://archive-gw-1.kat.ac.za/public/tools/continuum_sensitivity_calculator.html}.}. This sensitivity makes MeerKAT a powerful tool for observing a large number of targets in a reasonable amount of telescope time. However, short integration times typically lead to poor \emph{uv}-coverage and subsequently noisy point spread functions (PSFs). PSFs with complicated or bright sidelobes often create unwanted artefacts during image reconstruction, limiting the dynamic range. 

To mitigate the negative effects of short integration times, the total on-source time for each target was broken up into 12--minute scans, each separated by approximately 1\,hour. Targets with similar RA were grouped into epochs (or schedule blocks), with each epoch having a maximum RA range of 4\,hours. There are three epochs in this pilot study, labelled A, B, and C, the groupings of which are listed in Table \ref{tab:sampleobs}. Within a schedule block, scans of different targets were interleaved with each other, with a 2--minute visit to an appropriate phase calibrator after each target scan. By grouping targets into RA-constrained schedule blocks, several targets could share a phase calibrator, reducing the calibration overhead for that epoch. As a flux and bandpass calibrator, PKS\,J1939$-$6342 or PMN\,J0408$-$6545 was observed for ten minutes at the beginning of each epoch, and again every two hours. Appendix \ref{app:uv} shows the \emph{uv}-coverage from the final calibrated data from two target observations, with the effect of the chosen observing strategy clearly evident.

\section{Data Reduction}
\label{sec:processing}

We made use of the \textsc{oxkat\, v1.0}\footnote{\url{https://github.com/IanHeywood/oxkat}} software \citep{oxkatsoftware} to reduce and process the data for this project, with each epoch being processed separately. \textsc{oxkat} is a semi-automated Python-based reduction pipeline for MeerKAT data, currently optimised for L-band {\color{black}continuum} observations. The reduction procedure makes use of several radio astronomy software packages, including \textsc{casa} \citep{CASA} for cross-calibration and \textsc{wsclean}\footnote{\url{https://sourceforge.net/p/wsclean/wiki/Home/}} \citep{WSCLEAN} for general imaging. 

The \textsc{oxkat} pipeline is currently automated up to {\color{black}direction-independent self-calibration (referred to as second generation calibration or 2GC)} and imaging, with the {\color{black}cross-calibration (or first generation calibration, 1GC)} and initial imaging tasks run separately from the 2GC steps. Third generation calibration (3GC), which implements direction-dependent corrections to the visibilities, is not automated, but still possible within the \textsc{oxkat} framework. Here we detail the {\color{black}1GC, 2GC, and 3GC processing steps for our data,} implemented using the \textsc{oxkat} software. The {\color{black}total percentages of flagged data, mostly due to satellites and other radio frequency interference (RFI), and the final properties of the primary beam-corrected images} are listed in Table \ref{tab:sampleobs}.

\subsection{1GC: Cross-calibration and initial imaging}
All cross-calibration tasks, {\color{black}where known calibrator sources are used to correct the target phases and amplitudes,} are carried out in \textsc{casa}. The data are first averaged to 1024 channels and the field list is interrogated to determine primary calibrator, secondary calibrator, and target fields, as well as the target-secondary pairings. Initial flagging includes the known {\color{black}RFI-corrupted} frequency bands (amounting to $\sim$\,40\,per\,cent of the original bandwidth, dominated by RFI from satellites) as well as any poorly performing antennas. This step is then followed by automatic flagging of the calibrators, before models of the secondary calibrator fields are created. Both the secondary and primary calibrator models are then used to correct the target fields, after which the calibrated target fields are split into individual measurement sets. 

The corrected target data undergo automatic flagging before imaging with \textsc{wsclean}, using the multi-frequency and auto-masking options \citep{WSCLEANMFS} and a Briggs weighting of -0.3. {\color{black}With a synthesised beam of $\sim$\,8\,arcsec, this weighting provides the best compromise between angular resolution and noise sensitivity for our data. To capture all sources visible within MeerKAT's field of view, we image a 3.1{\degree} $\times$ 3.1{\degree} region centred on the cluster target. This is $\sim20$\,per\,cent wider than the MeerKAT primary beam at the lowest L-band frequency. We also note that the effective observing frequency, $\nu_0$, of an image is dependent on the final flagged bandwidth, which varies across the three epochs.} 

\subsection{2GC: Self-calibration}
Self-calibration of the target visibilities, {\color{black}where gain solutions derived using only target data are used to iteratively improve its calibration,} is carried out using \textsc{casa} and \textsc{wsclean}. First, the initial mask from the 1GC target imaging step is used to re-image the corrected data: {\color{black}the mask tells the software where to look for true emission and an accurate mask can greatly improve image quality. The mask is updated during re-imaging, after which a target sky model is predicted}. The sky model is then used in \textsc{casa} to self-calibrate the target data, {\color{black}which determines a set of single-direction calibration solutions based on the phase centre of the dataset}. Finally, the 2GC-corrected data are re-imaged and the mask and sky model updated.

For most of our targets, 2GC improves the image quality {\color{black}by reducing artefacts} and lowers the noise floor by {\color{black} 1.3--20.3 per cent, depending on the field,} resulting in {\color{black}image qualities} sufficient for our scientific purposes. However, in four of our 13 targets, namely J0019.6$+$0336, J0034.4$+$0225, J0240.0$+$0115, and J0248.2$+$0238, 2GC processing results in a noisier image {\color{black}with larger artefacts}. In each of the four cases, the field is populated by several bright ($>100$\,mJy), extended sources, spread out over the $\sim$\,9\,deg$^2$ imaging field. In these cases, applying a {\color{black}calibration solution across the full field of view based only on the phase centre direction} is insufficient and exacerbates amplitude errors around the brightest sources. For these fields, the measurement set is therefore reverted to the 1GC stage before carrying out direction-dependent corrections.

\subsection{3GC: Direction-dependent calibration}
For the four fields {\color{black}where 2GC processing resulted in a poorer quality image,} the \textsc{ddfacet}\footnote{\url{https://github.com/saopicc/DDFacet}} \citep{DDFACET} and \textsc{killms}\footnote{\url{https://github.com/saopicc/killMS}} \citep{KILLMS2014, KILLMS2015} packages are used to {\color{black}improve results through direction-dependent calibration. This is often needed for wide-field imaging where phase and amplitude gains may vary significantly across the field of view. With \textsc{ddfacet} and \textsc{killms}, the imaging field is broken up into several regions, called facets, each with its own phase centre, typically a bright source. Calibration solutions are determined for each facet separately before being applied to the data. For our four datasets, the} 1GC-corrected visibilities are re-imaged with \textsc{ddfacet}, using between six and ten bright sources (depending on the field) to determine the facets. \textsc{killms} uses these facets and source positions to determine phase and amplitude corrections on a per-facet basis. A second round of imaging in \textsc{ddfacet} applies these corrections on the fly to produce a 3GC image {\color{black}with reduced artefacts around bright sources.} After 3GC, the ACT-CL\,J0034.4$+$0225 field is still affected by direction-dependent artefacts. This field is particularly complex, with several bright (10--65\,mJy), extended sources within the primary beam.

{\color{black}Additional calibration was also required for the J0040.8$-$4407 field, which} is contaminated by a 2.6\,Jy resolved source, $\sim$\,15\,arcmin from the pointing centre. {\color{black}Self-calibration improved the central rms noise by 20\,per\,cent, however the bright source was still a significant contaminant, with a central 2GC noise floor of 100\,$\mu$Jy\,beam\per. To remove this bright source from the visibilities,} we peeled the bright source using the \textsc{cubical}\footnote{\url{https://github.com/ratt-ru/CubiCal}} package \citep{CUBICAL}. The peeling process {\color{black}uses calibration solutions towards the direction of the interfering source to remove the source contributions from the visibilities before re-imaging. After peeling, the central rms noise level is improved} by 69\,per\,cent to 31.8\,$\mu$Jy\,beam\per, and the peak {\color{black}image brightness is reduced} to 0.11\,Jy\,beam\per. Additional processing of this field will be necessary to further reduce residual artefacts radiating from the peeled source region. 

\subsection{Primary beam corrections}
Once a final image has been obtained, we use the \textsc{katbeam}\footnote{\url{https://github.com/ska-sa/katbeam}} package to create primary beam corrected images for analysis. The final images are masked such that they contain only regions where the primary beam {\color{black}response remains} above 30\,per\,cent of the value at the phase centre. Primary beam correction increases the central rms noise by a median of 1.3 per cent across our 13 fields.

\begin{figure*}
    \includegraphics[width=0.33\textwidth,clip=True,trim=0 0 0 0]{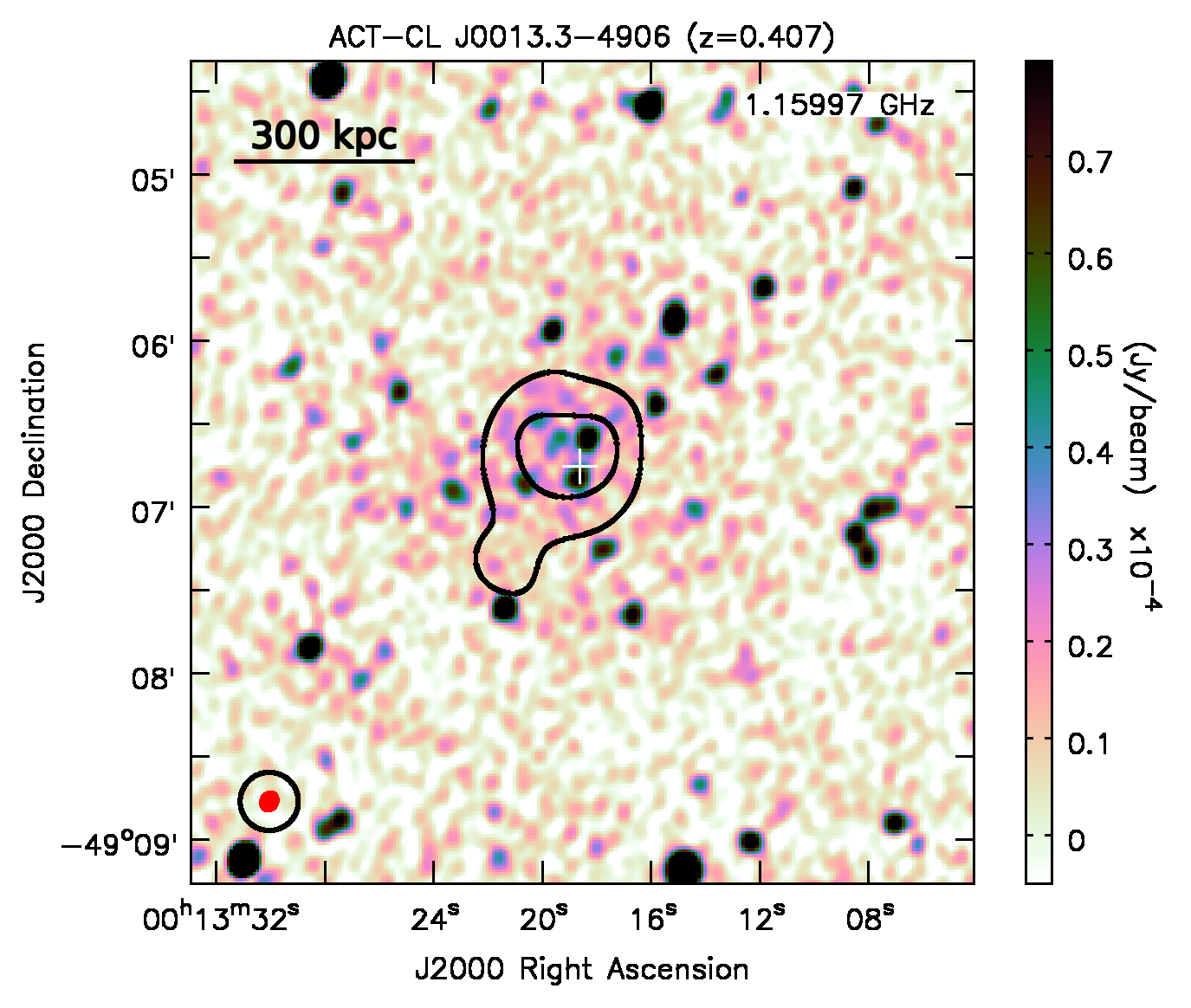}
    \includegraphics[width=0.33\textwidth,clip=True,trim=0 0 0 0]{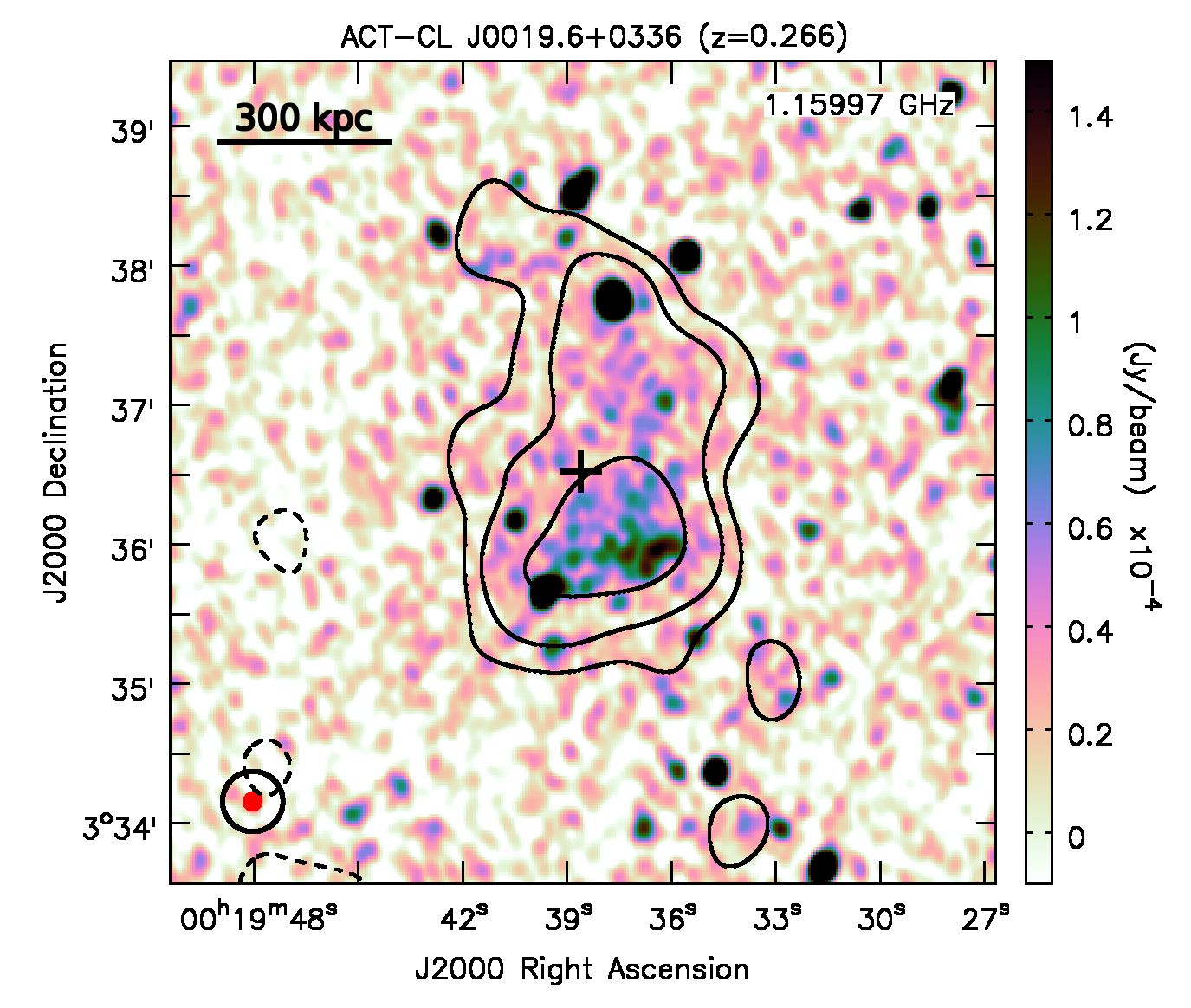}
    \includegraphics[width=0.33\textwidth,clip=True,trim=0 0 0 0]{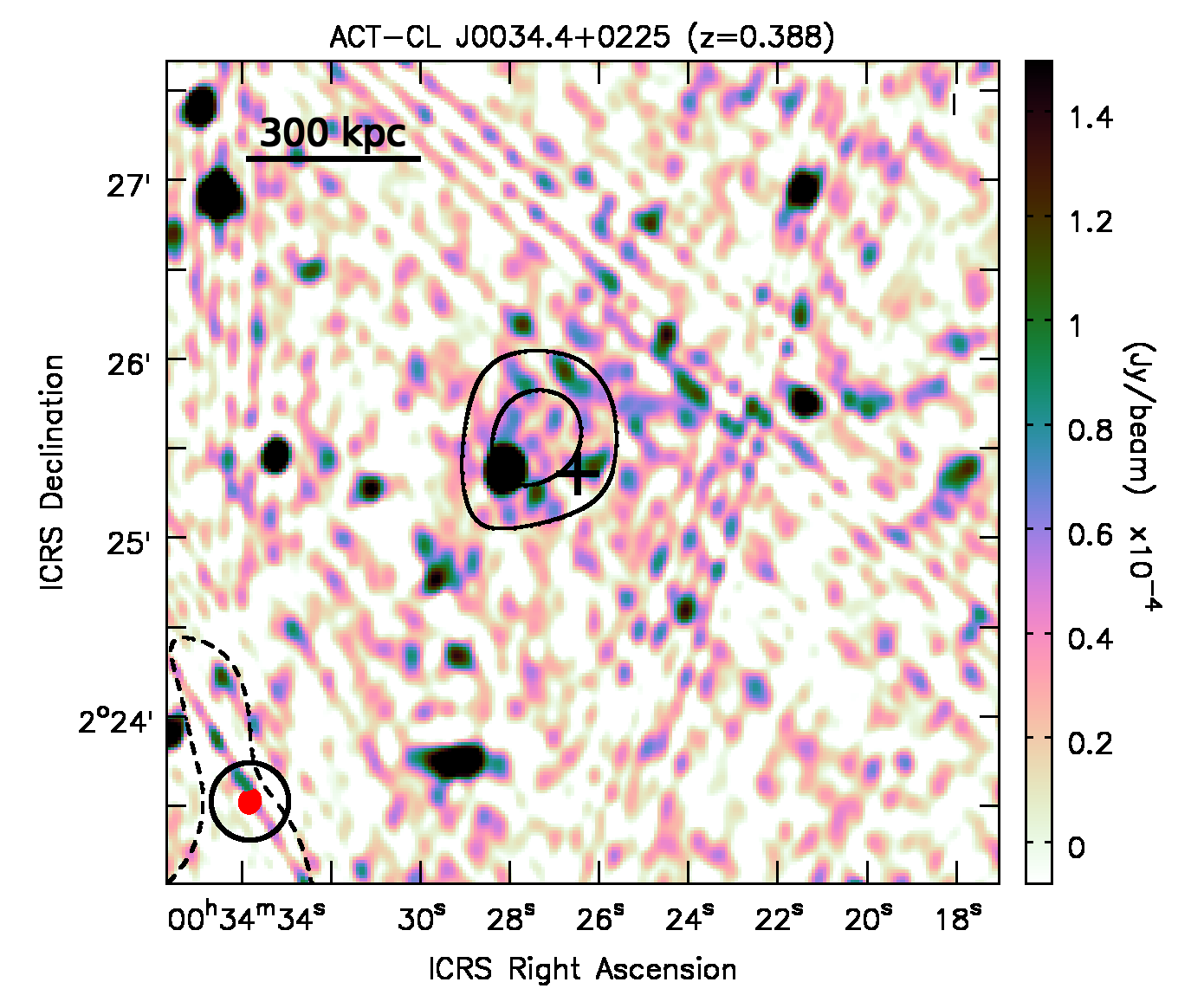}\\
    \includegraphics[width=0.33\textwidth,clip=True,trim=0 0 0 0]{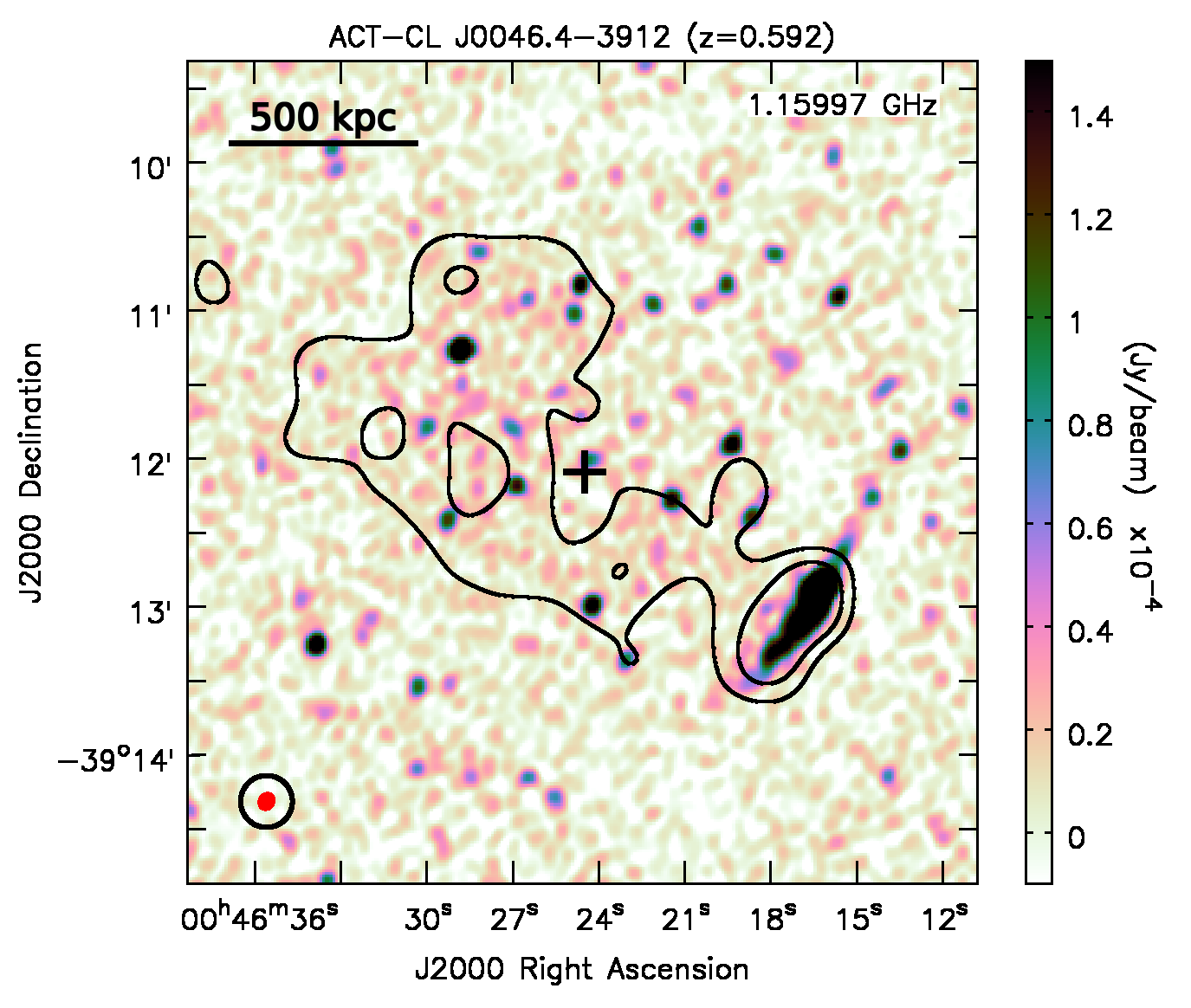}
    \includegraphics[width=0.33\textwidth,clip=True,trim=0 0 0 0]{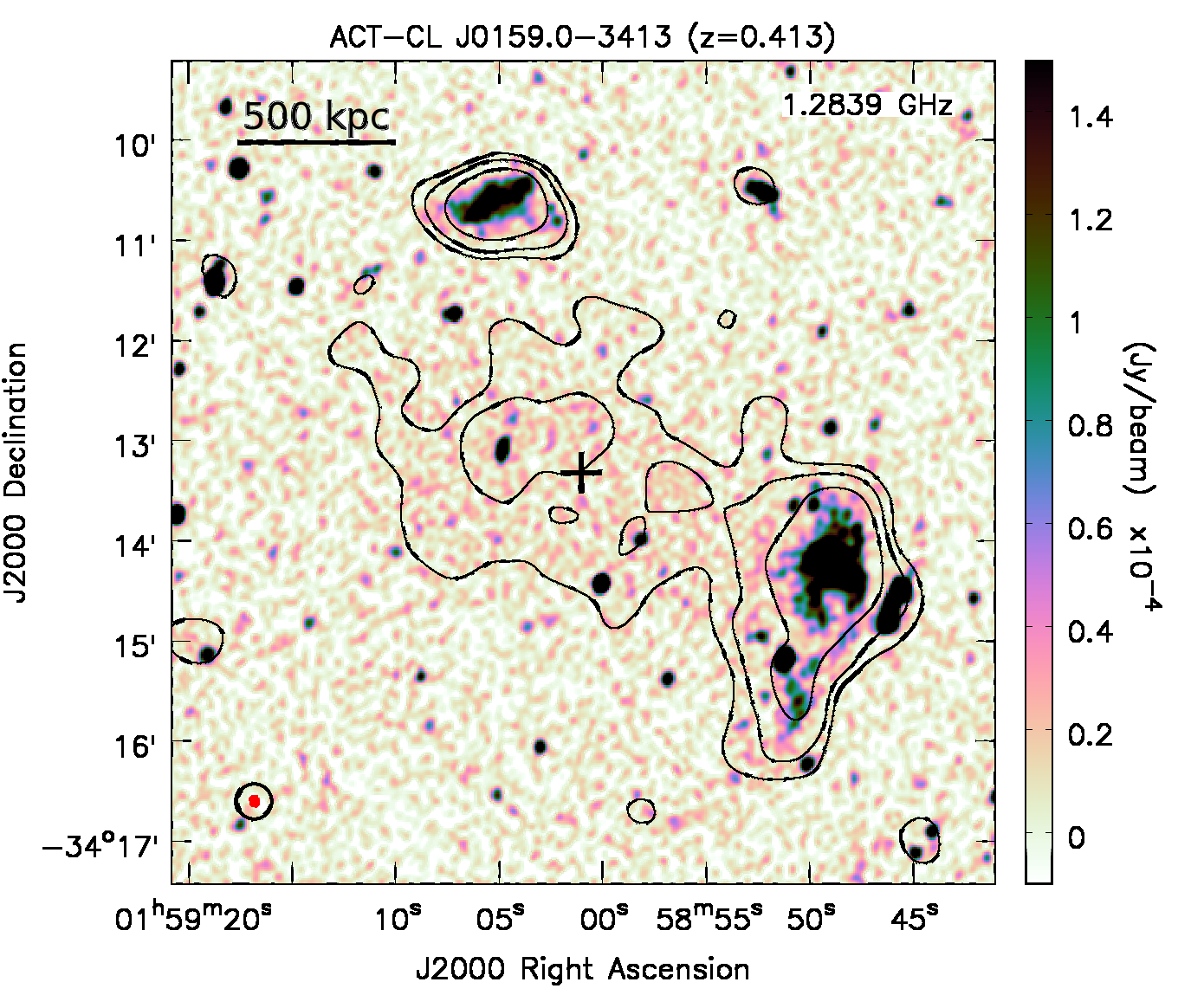}
    \includegraphics[width=0.33\textwidth,clip=True,trim=0 0 0 0]{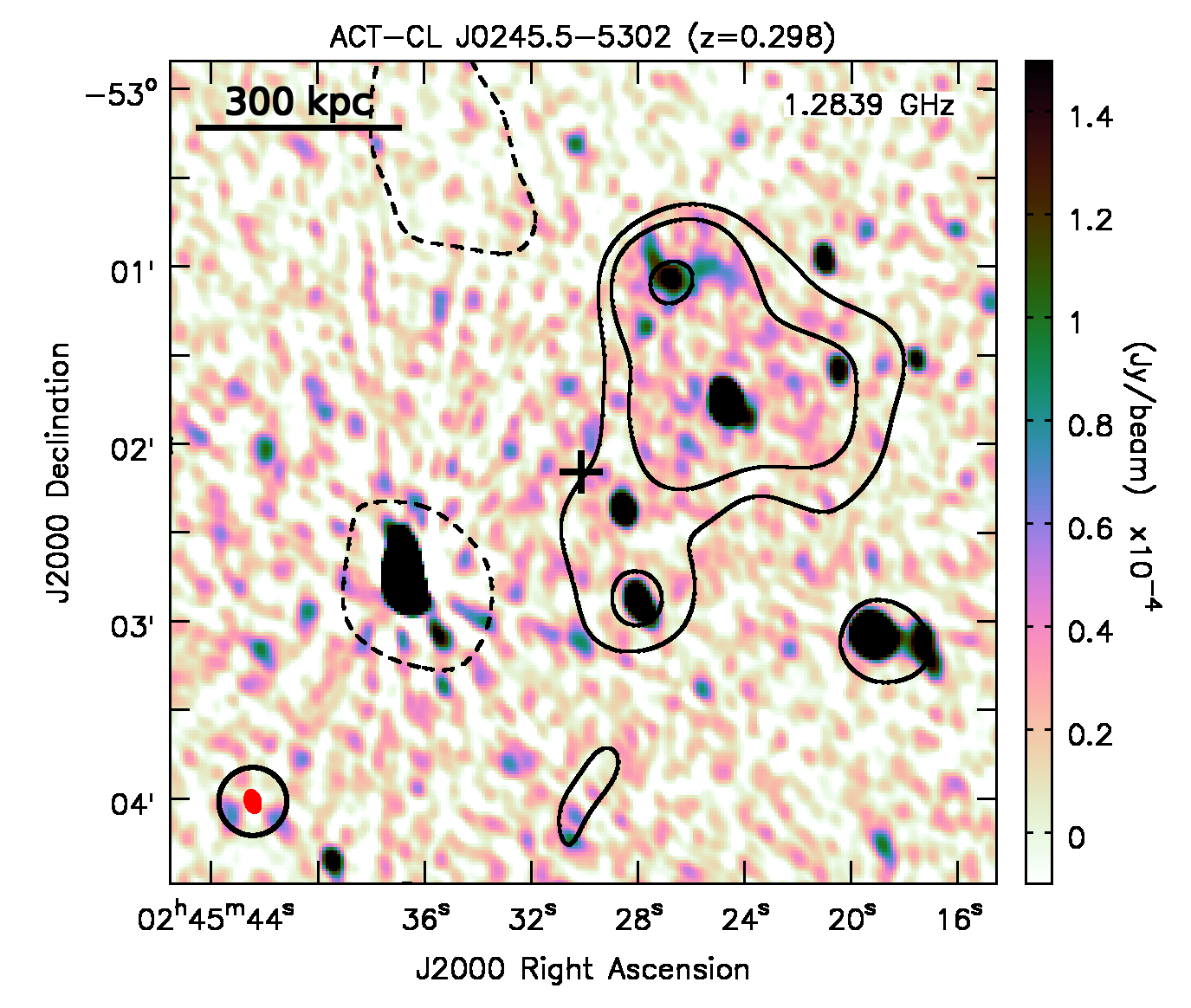}\\
    \includegraphics[width=0.33\textwidth,clip=True,trim=0 0 0 0]{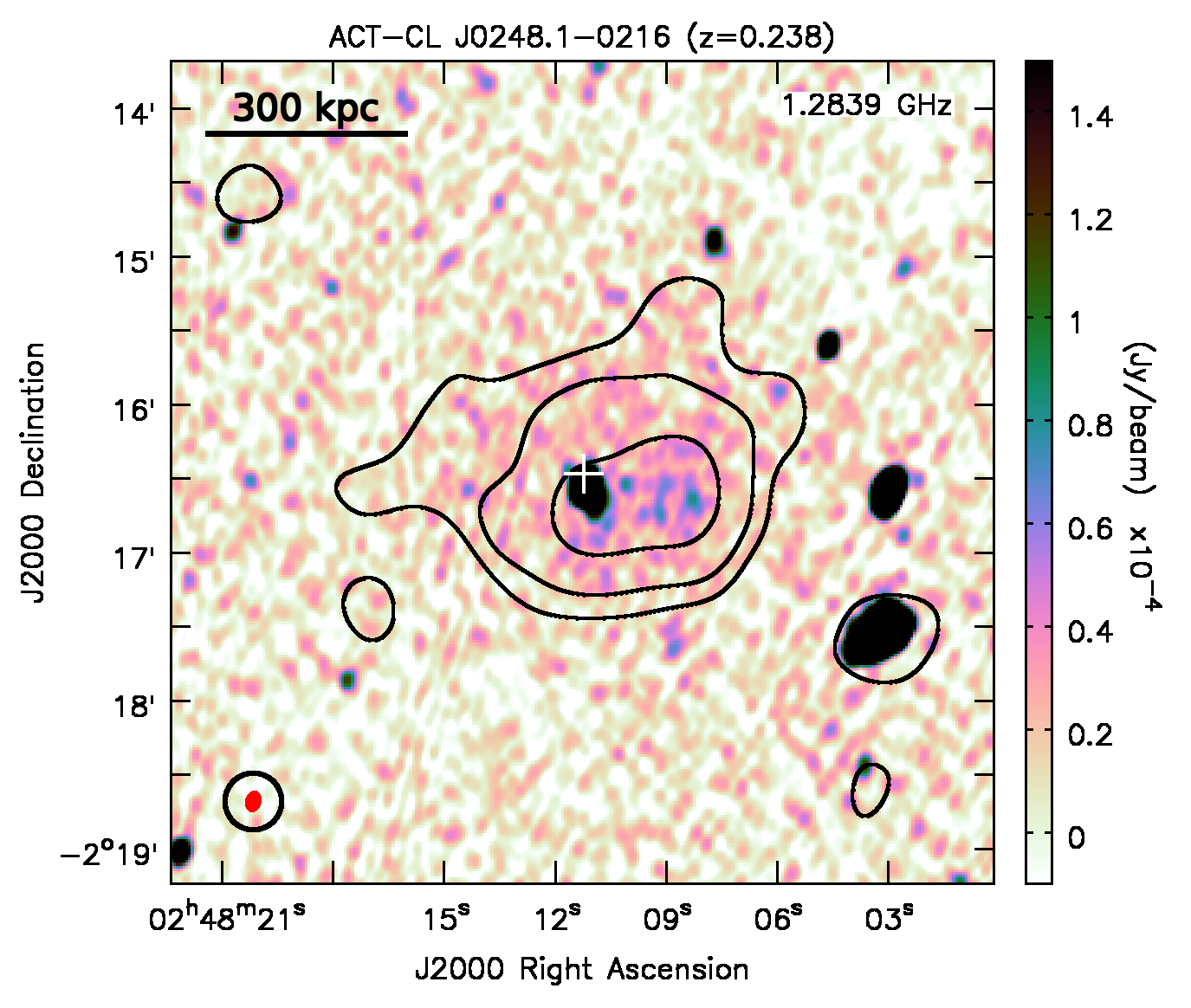}
    \includegraphics[width=0.33\textwidth,clip=True,trim=0 0 0 0]{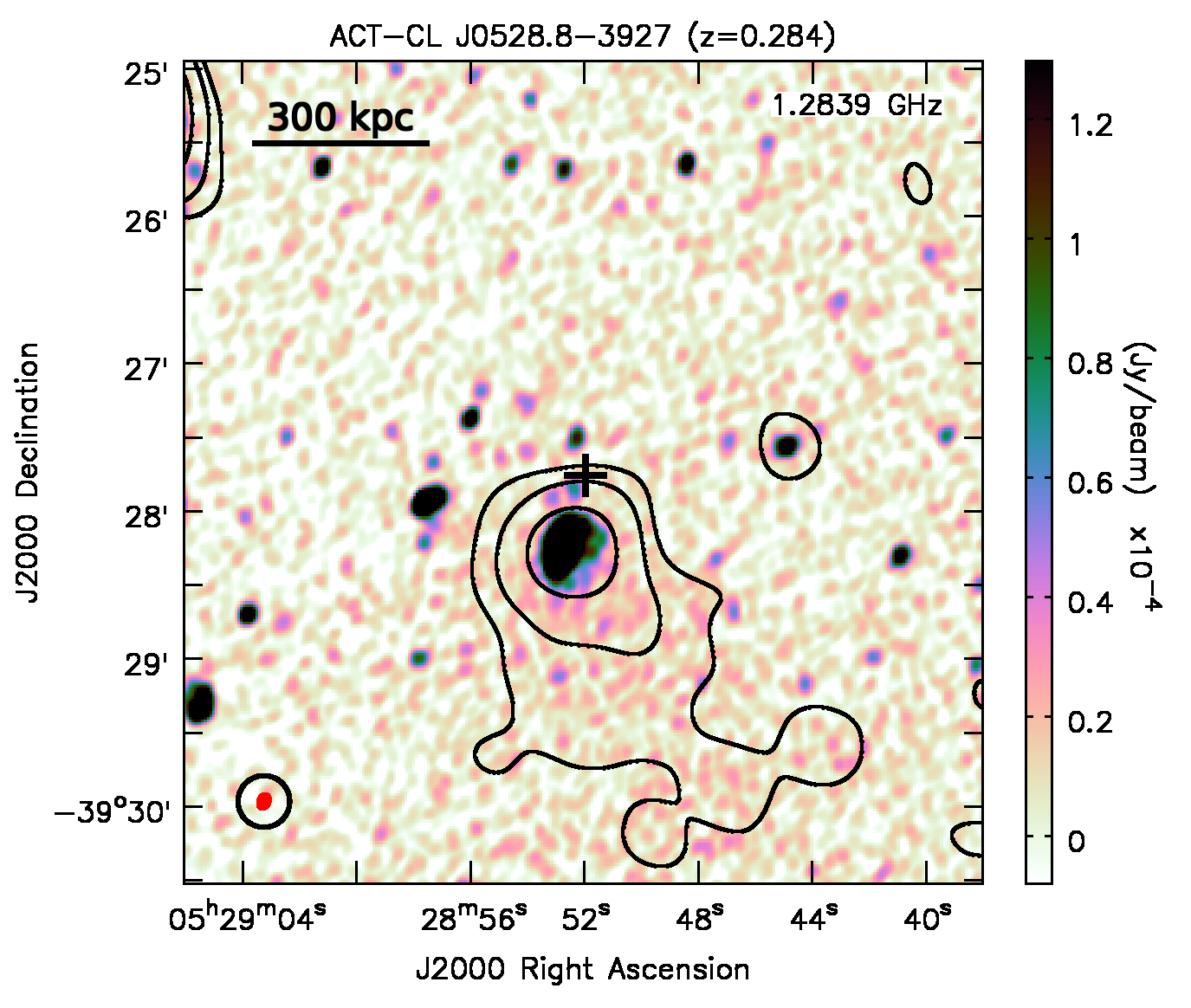}
    \includegraphics[width=0.33\textwidth,clip=True,trim=0 0 0 0]{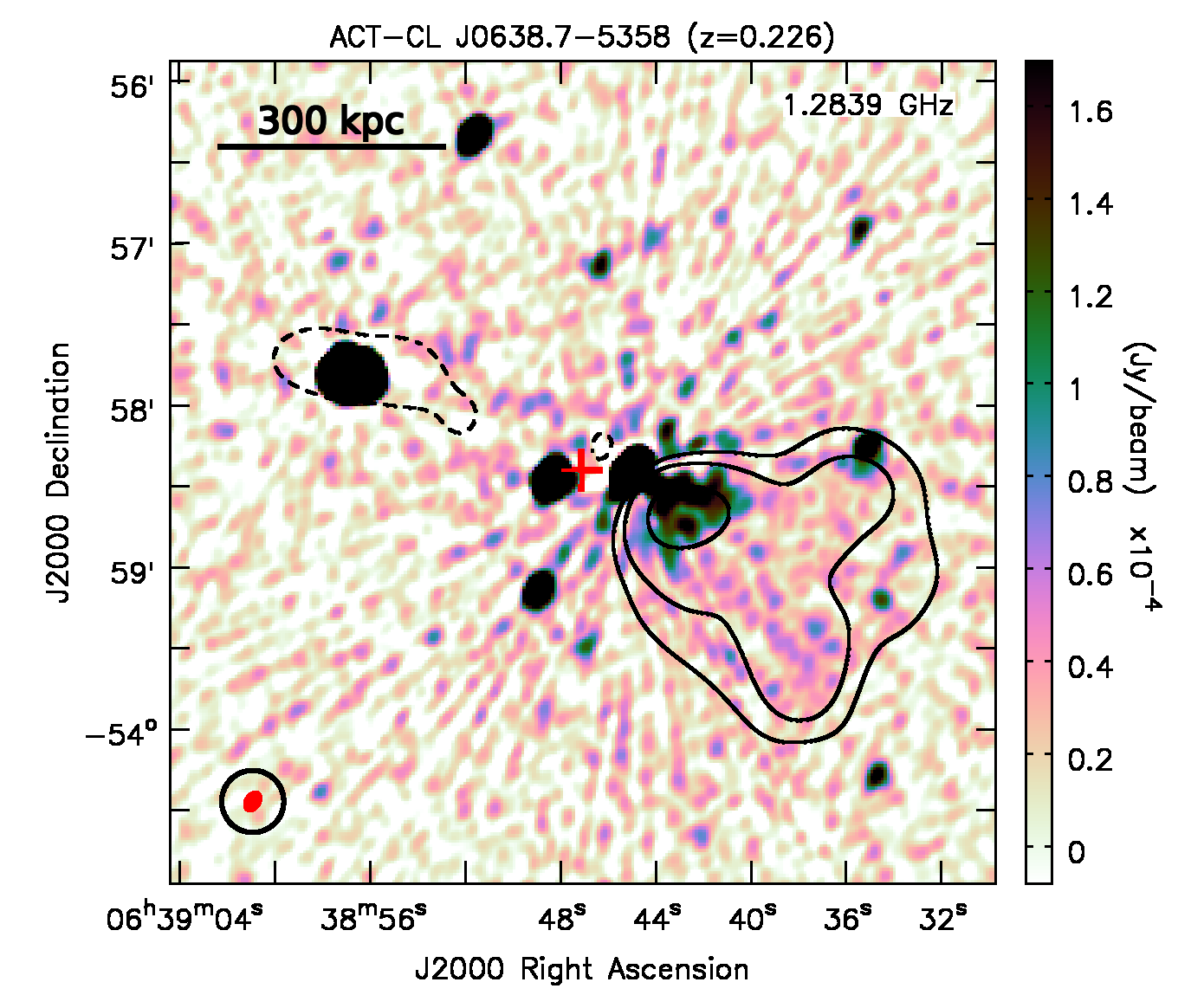}
    \caption{Full-resolution MeerKAT L-band images for the nine clusters with diffuse emission detections in our sample, with low resolution contours from the LS map overlaid. In all panels, contours are at [-3,\,3,\,5,\,10]\,$\times\,\sigma_{\rm LS}$. Negative contours are dashed. The synthesised beam for both the full-resolution (filled red ellipse) and LS maps (open black circle) are indicated in the lower left of each panel. The beam sizes and central rms noise are provided in Table \ref{tab:sampleobs}. The physical scale at the cluster redshift is indicated in the upper left of each panel, and the cross indicates the position of the ACT SZ peak. Artefacts remain in the ACT-CL\,J0034$+$0225 (top right) and ACT-CL\,J0638.7$-$5358 (bottom right) cluster fields, however we are still able to recover extended emission in the cluster region. }
    \label{fig:gallery}
\end{figure*}

{\renewcommand{\arraystretch}{1.5}
\begin{table*}
	\centering
	\caption{Measured properties of all diffuse emission detections in the sample. Cols: (1) ACT DR5 cluster name; (2) Cluster redshift; (3) ACT SZ weak-lensing-calibrated mass; \changes{(4) ACT SZ uncalibrated mass;} (5) Largest angular size, in arcmin, of diffuse emission; (6) Largest physical size, in Mpc, of diffuse emission at the cluster redshift; (7) Classification: Halo (H), Relic (R), candidate (c); (8) New detection (or not); (9) Effective observed frequency; (10) Integrated flux density in mJy; \changes{(11) $k$-corrected radio power at 1.4\,GHz, assuming $\alpha = -1.3 \pm 0.4$;} (12) Signal-to-noise of the MeerKAT detection; (13) Comments.}
	\label{tab:results}
	\begin{tabular}{lc..cccccrc.l} 
		\hline
		\multicolumn{1}{c}{(1)} & {(2)} & \multicolumn{1}{c}{(3)} & \multicolumn{1}{c}{(4)} & (5) & \multicolumn{1}{c}{(6)} & (7) & (8) & (9) & \multicolumn{1}{c}{(10)} & \multicolumn{1}{c}{(11)} & \multicolumn{1}{c}{(12)} & \multicolumn{1}{l}{(13)}\\
		Name & z & \multicolumn{1}{c}{$M_{500}$} & \multicolumn{1}{c}{$M_{500}^{\rm unc}$} & LAS & \multicolumn{1}{c}{LLS} & Class & New? & $\nu_0$ & \multicolumn{1}{c}{$S_{\nu_0}$}& \multicolumn{1}{c}{log$P_{\rm 1.4GHz}$} & \multicolumn{1}{c}{SNR} & Comments \\
		(ACT-CL...) &  & \multicolumn{2}{c}{($10^{14}\,M_\odot$)} & (\arcmin) & \multicolumn{1}{c}{(Mpc)} & & & (GHz) & \multicolumn{1}{c}{(mJy)}& \multicolumn{1}{c}{(W\,Hz$^{-1}$)} & \\
		\hline 
		J0013.3$-$4906 &  0.407 &  6.8 & 5.4 & 0.9 & 0.29 & cH & \checkmark & 1.16 &  0.54\,$\pm$\,0.08 & 23.44$^{+0.08}_{-0.11}$ & 6.7 \\ 
		J0019.6$+$0336 &  0.266 & 10.2 & 8.3 & 3.3 & 0.81 &  H & \checkmark & 1.16 &  9.16\,$\pm$\,0.57 & 24.13$^{+0.06}_{-0.06}$ & 16.1\\ 
		J0034.4$+$0225 &  0.388 & 8.1 & 6.5 & 1.1 & 0.35 & cH & \checkmark & 1.16 &  1.26\,$\pm$\,0.20 & 23.76$^{+0.09}_{-0.11}$ & 6.4 \\ 
		J0046.4$-$3912 &  0.592 &  7.9 & 6.2 & 1.4 & 0.56 &  R & \checkmark & 1.16 &  3.48\,$\pm$\,0.19 & 24.66$^{+0.08}_{-0.10}$ & 18.7 & 1.8{\arcmin} SW of centre. \\ 
		               &        &      &     & 3.2 & 1.28 & H & \checkmark & 1.16 &  3.58\,$\pm$\,0.27 & 24.67$^{+0.08}_{-0.10}$ & 13.3 \\ 
		J0159.0$-$3413 &  0.413 &  9.1 & 7.2 & 3.3 & 1.08 & R & \checkmark & 1.28 & 13.99\,$\pm$\,0.73 & 24.92$^{+0.06}_{-0.07}$ & 19.1 & 3.0{\arcmin} SW of centre. \\ 
		               &        &      &     & 1.7 & 0.56 &  R & \checkmark & 1.28 &  4.25\,$\pm$\,0.23 & 24.41$^{+0.06}_{-0.07}$ & 18.6 & 2.8{\arcmin} N of centre. \\ 
		               &        &      &     & 3.7 & 1.22 & H & \checkmark & 1.28 &  9.71\,$\pm$\,0.57 & 24.77$^{+0.06}_{-0.07}$ & 17.0 \\ 
		J0245.5$-$5302 &  0.298 & 10.7 & 8.7 & 2.0 & 0.53 &  H & \checkmark & 1.28 &  3.82\,$\pm$\,0.28 & 24.02$^{+0.05}_{-0.06}$ & 13.7 \\ 
		J0248.1$-$0216 &  0.238 & 9.9 & 8.0 & 2.9 & 0.66 &  H & \checkmark & 1.28 &  6.46\,$\pm$\,0.41 & 24.02$^{+0.05}_{-0.05}$ & 15.9 \\ %
		J0528.8$-$3927 &  0.284 &  9.0 & 7.2 & 2.1 & 0.54 &  H & \checkmark & 1.28 &  3.58\,$\pm$\,0.26 & 23.94$^{+0.05}_{-0.06}$ & 13.5 & Extends SW from BCG \\ 
		J0638.7$-$5358 &  0.226 & 12.5 & 10.5 & 2.2 & 0.48 &  H &            & 1.28 &  4.68\,$\pm$\,0.34 & 23.83$^{+0.05}_{-0.05}$ & 13.8 & \citet{ASKAPJ0638} \\ 
		\hline
	\end{tabular}
\end{table*}
}
\section{Results}
\label{sec:results}

Table \ref{tab:sampleobs} lists the central rms noise, $\sigma_{\rm FR}$, and synthesised beam, $\theta_{\rm synth,FR}$, for the final {\color{black}full-resolution, primary beam-corrected} images. The native resolutions are in the range 6.7--8.3\,arcsec, with cluster region noise levels between 7.5 and 31.8\,$\mu$Jy\,beam$^{-1}$. 

To increase the sensitivity to faint extended emission, we need to image at lower resolution. However, extended structure would be contaminated by source blending if compact sources were not first removed. This step is typically taken using
{\color{black}$uv$-range restricted} source modeling and model-subtraction from the {\color{black}visibilities}. This method failed on our datasets due to the combination of short integration times and MeerKAT's dense core: {\color{black}restricting the model to physical scales less than 100\,kpc\footnote{The smallest types of diffuse emission are typically 100--400\,kpc across.}
our source model had too little data to accurately model the compact source flux; reducing the $uv$-cut to allow enough data for accurate flux characterisation led to models which, when removed, led to over-subtraction of the extended emission}. To image extended emission without compact sources, we instead implemented the image-plane filtering technique described in \citet{LBWanFilter}, whereby emission on scales 1--3\,times the synthesised beam is filtered out, {\color{black}creating an image with only the larger scale emission}. {\color{black}As no convolution takes place, the large-scale} image has the same units and resolution as the original, but has a negatively offset zero-level that needs to be corrected before any quantitative analysis. After correction for this zero-level, the image is convolved, for aesthetics, with a Gaussian slightly larger than the filter size. To verify the success of the compact source removal, we checked the filtered maps (before convolving) at the positions of isolated compact sources and find values consistent with noise. The rms noise, $\sigma_{\rm LS}$, and convolved beam size, $\theta_{\rm synth,LS}$, of the smoothed {\color{black}large-scale-emission} images (hereafter the LS images) are given in the last two columns of Table \ref{tab:sampleobs}. 

Figure \ref{fig:gallery} shows the cluster regions for the nine systems in which we detect diffuse extended emission {\color{black}(SNR\,$>5$), with two of the nine clusters containing more than one diffuse source}. The full resolution image is shown in the colour map, with the synthesised beam indicated by the filled red ellipse in the lower left of each panel. Black contours show the $[-3,\,3,\,5,\,10]\,\times\,\sigma_{\rm LS}$ levels from the relevant LS map. Negative contours are dashed, and the LS synthesised beam is shown by the bold black ellipse in the lower left of each panel. The cross indicates the position of the ACT SZ peak. The physical scale at the cluster redshift is indicated in the upper left of each panel. Figure \ref{fig:nondets} in the appendix shows the cluster regions for the four targets with no diffuse emission detection. 

We detect a $\sim$\,560\,kpc radio relic in ACT-CL\,J0046.4$-$3912, and double relics in ACT-CL\,J0159.0$-$3413. Sub-Mpc radio halos are detected in five other systems, with candidate halos in ACT-CL\,J0013.3$-$4906 and ACT-CL\,J0034.4$+$0225. Given the observed sizes, the candidate halos could also be mini-halos, although unlikely if these are strongly dynamically disturbed systems. An X-ray study of these systems may assist in a firmer classification. The LS maps confirm the full-resolution detections in all nine systems, and reveal giant radio halos (physical size\,$>$\,1\,Mpc) in ACT-CL\,J0046.4$-$3912 and ACT-CL\,J0159.0$-$3413. Table \ref{tab:results} lists the largest angular size (LAS), largest projected physical size (LLS) at the cluster redshift, classification, effective observing frequency $\nu_0$, measured flux density $S_{\nu_0}$, and SNR for each diffuse emission detection. Flux densities are measured within the 3$\sigma$ region and we assume a 5\,per\,cent MeerKAT amplitude uncertainty. The total flux density uncertainty ${\Delta}S$ is measured as ${\Delta}S = \sqrt{\left(0.05 S\right)^2 + N \sigma^2},$
where $N$ is the number of beams within the 3$\sigma$ region. With the exception of the radio relic in ACT-CL\,J0046.4$-$3912 and the northern relic in ACT-CL J0159.0$-$3413, which do not have embedded compact sources, all flux density measurements were made in the LS image. For the two former cases, the flux density is measured in the full resolution image.

\section{Discussion}

All but one of our detections are new, with the halo in ACT-CL\,J0638.7$-$5358 having been detected by ASKAP \citep{ASKAPJ0638}. We do not detect the full ASKAP halo emission in this system; however, we do resolve a peaked region in the halo that is morphologically aligned with the X-ray emission of the infalling subcluster seen in figure 6 from \citet{ASKAPJ0638}.

\changes{To compare our detections to the literature, we determine $k$-corrected 1.4\,GHz radio powers, $P_{\rm1.4 GHz}$, for all diffuse emission detections. To extrapolate the MeerKAT flux densities to 1.4\,GHz we assume a fiducial spectral index of $\alpha\,=\,-1.3\,\pm\,0.4$, adopting the spectral power law convention of $S_\nu\,\propto\,\nu^\alpha$. The choice of spectral index uncertainty allows for the wide range of observed spectral indices for radio halos and relics. We note that the $k$-corrected radio powers will be higher if the detected sources have very steep spectra ($\alpha < -1.7$), however, as the MeerKAT reference frequency is quite close to 1.4\,GHz, the effect will not be large. 
}

\begin{figure*}
    \centering
    \includegraphics[width=0.47\textwidth,clip=True,trim=10 30 50 30]{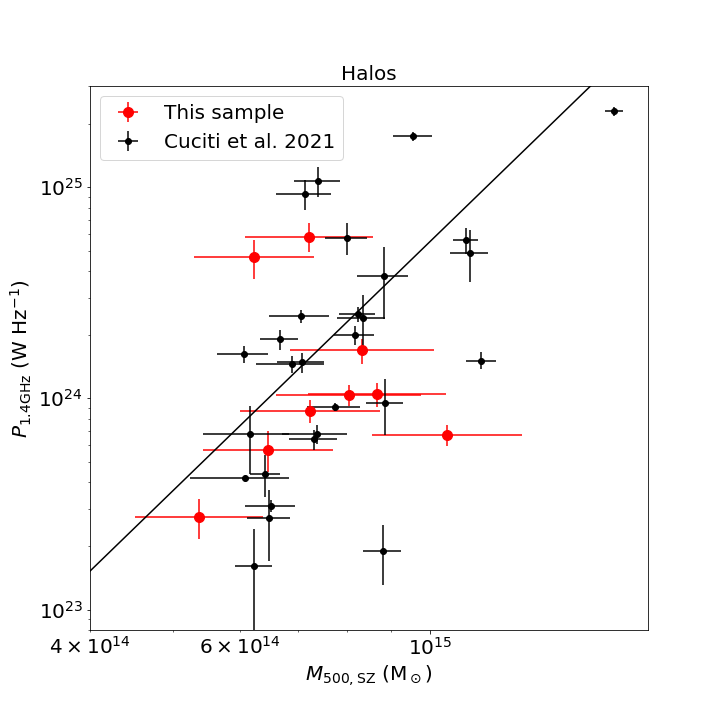}
    \includegraphics[width=0.47\textwidth,clip=True,trim=10 30 50 30]{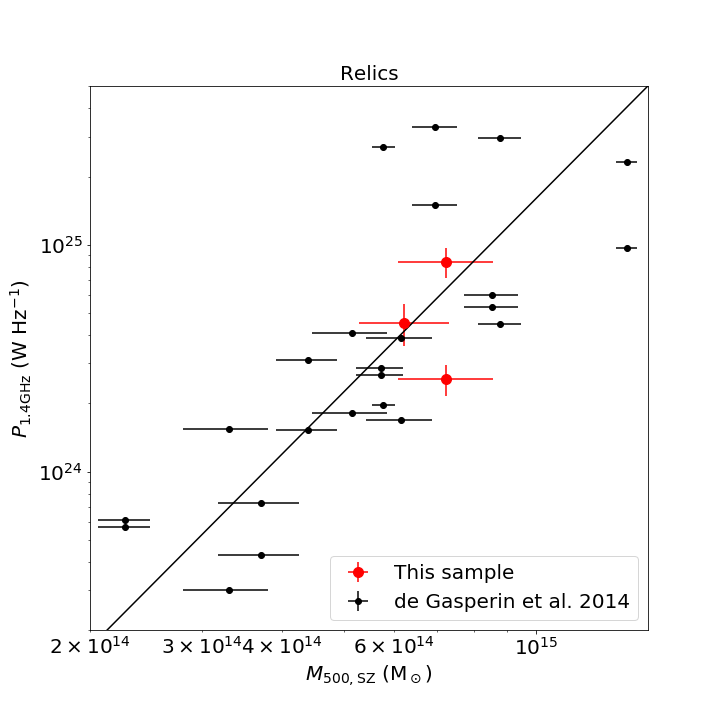}
    \caption{\changes{1.4\,GHz radio power versus SZ M$_{500}$ mass for the diffuse emission detections listed in Table \ref{tab:results} (large red points), compared to literature values (small black points). The literature values are from \textit{Planck} \citep{Planck2014,Planck2016} and we therefore use the uncorrected SZ masses from ACT DR5 to accurately compare our results. \textbf{Left:} Radio halos in our sample, including candidates, compared with radio halos from the statistical sample in \citet{2021A&A...647A..51C}, where the solid line indicates their BCES Y|X best fit. \textbf{Right:} Radio relics in our sample compared with literature values from \citet{FdG2014}, with the solid line indicating their ``double + single relics'' best fit.} }
    \label{fig:powerplots}
\end{figure*}

\changes{The calculated 1.4\,GHz radio powers for all detections are provided in column 11 in Table \ref{tab:results}, and we show their comparison with known scaling relations (\citealt{2021A&A...647A..51C} for radio halos; \citealt{FdG2014} for relics) in Figure \ref{fig:powerplots}. We note that the literature results use SZ masses from \textit{Planck} which have not been calibrated against weak lensing results, and are therefore systematically lower than our ACT DR5 masses used in selection. In Figure \ref{fig:powerplots} we therefore use the `uncorrected' ACT DR5 SZ masses\footnote{\changes{The `uncorrected' ACT masses are referred to as $M_{\rm 500c}^{\rm Unc}$ in \citet{2021ApJS..253....3H} and are in the \textsc{M500cUncorr} columns of the ACT DR5 catalogue available on LAMBDA (\url{https://lambda.gsfc.nasa.gov/product/act/actpol_prod_table.cfm}).}} for our clusters, given in column 4 of Table \ref{tab:results}, to more accurately compare results. All of our detections lie within the scatter of the correlations, indicating that we successfully removed contaminating compact emission. We note that our ACT-CL\,J0638.7$-$5358 radio halo, residing in the most massive of our clusters, lies at the edge of the scatter in the literature values, and in the region typically associated with ultra-steep spectrum sources. However, our power is likely underestimated due to missing flux, as described above, with \citet{ASKAPJ0638} quoting a halo power a factor of $\sim 2.8$ higher, which would move it closer to the correlation.
}

\section{Conclusions}
\label{sec:conclusion}
The MERGHERS survey is a planned project to observe a statistically significant number ($\sim$\,200) of galaxy clusters with MeerKAT, in order to probe the cosmic and mass evolution of diffuse cluster radio emission and cosmic magnetic fields. Observing a statistically significant number of clusters in a single observing season requires relatively short ($\lesssim\,1$\,hour) on-source times. We have presented the results of a MERGHERS pilot study which validates this as feasible, confirming MeerKAT's suitability for large cluster studies at L-band.

In this work, we have carried out pilot observations of 13 SZ-selected galaxy clusters detected in the ACT DR5 maps, observing each cluster for less than two hours, with most being observed for only 24 minutes. Short observations can be susceptible to poor \emph{uv}-coverage, which in turn negatively affects the point spread function and subsequently limits image quality. Our chosen observing strategy, to be implemented in the full MERGHERS programme, mitigates this risk by splitting a target's observation time into several smaller chunks, interspersed with observations of other clusters, so that a wider range of hour angle is covered for every target.

By mitigating sparse \emph{uv}-coverage, and employing a range of modern data reduction and imaging techniques, we reached or improved upon the predicted rms noise floor, based on our total time-on-source, for most target fields. However, for the fields which underwent 3GC processing, our final rms noise is higher by a factor of 1.1--1.9 compared to the predicted value. This is likely because theoretical noise estimates do not take into account bright source contamination. For future observations, for fields with significant bright sources in all-sky surveys, full-field modelling may be a preferred method for estimating required source integration times. Due to its several extended bright sources, the ACT-CL\,J0034.4$+$0225 field is a good candidate for testing such algorithms, as well as testing more advanced reduction techniques.

We are able to reliably detect diffuse emission in 70 per cent of our clusters, with our highest redshift detection being at $z\,=\,0.592$. All but one of our detections are new and the full set, provided in Table \ref{tab:results}, can be summarised as follows: two systems host both radio relic(s) and a giant radio halo, five systems have radio halos, and two have candidate radio halos. \changes{Our estimated 1.4\,GHz radio powers for all detections, assuming a fiducial spectral index of $-1.3\,\pm\,0.4$, are consistent with known scaling relations. Determination of in-band spectral indices and power upper limits for the non-detections are} outside of the scope of this detection paper. When combined with multiwavelength data to verify the cluster dynamical state, these radio results will allow us to study cluster magnetic fields out to $z \sim 0.6$.

Our MERGHERS pilot project has shown that MeerKAT's L-band provides sufficient resolution and sensitivity to reliably separate diffuse emission from compact sources. As MeerKAT also operates at lower frequencies where the steep-spectrum diffuse emission is brighter, a similar test can be carried out with UHF-band data to determine the limiting redshift with the lower UHF resolution. 


\section*{Data Availability}
The data underlying this article will be shared on reasonable request to the corresponding author.

\section*{Acknowledgements}

The MeerKAT telescope is operated by the South African Radio Astronomy Observatory (SARAO), which is a facility of the National Research Foundation, an agency of the Department of Science and Innovation.

KK, DP, and SS acknowledge funding support from SARAO. JPH acknowledges support from NSF Astronomy and Astrophysics Research Program award number 1615657. MH and KM acknowledge support from the National Research Foundation of South Africa. CP acknowledges support by the European Research Council under ERC-CoG grant CRAGSMAN-646955. CS acknowledges support from the Agencia Nacional de Investigaci\'on y Desarrollo (ANID) through FONDECYT Iniciaci\'on grant no. 11191125. The research of OS is supported by the South African Research Chairs Initiative of the Department of Science and Technology and National Research Foundation. ZX is supported by the Gordon and Betty Moore Foundation.

This research made use of the Astropy,\footnote{http://www.astropy.org} \citep{astropy2013, astropy2018}, NumPy \citep{numpy2020}, and SciPy \citep{scipy2020} Python packages. Astropy is a community-developed core Python package for Astronomy. The Common Astronomy Software Applications (CASA) package is developed by an international consortium of scientists based at the National Radio Astronomical Observatory (NRAO), the European Southern Observatory (ESO), the National Astronomical Observatory of Japan (NAOJ), the Academia Sinica Institute of Astronomy and Astrophysics (ASIAA), the CSIRO division for Astronomy and Space Science (CASS), and the Netherlands Institute for Radio Astronomy (ASTRON) under the guidance of NRAO. The National Radio Astronomy Observatory is a facility of the National Science Foundation operated under cooperative agreement by Associated Universities, Inc.




\bibliographystyle{mnras}
\bibliography{mkatot} 



\appendix
\section{UV-coverage plots}\label{app:uv}
\begin{figure}
    \centering
    \includegraphics[width=0.98\columnwidth,clip=True,trim=1 0 0 10]{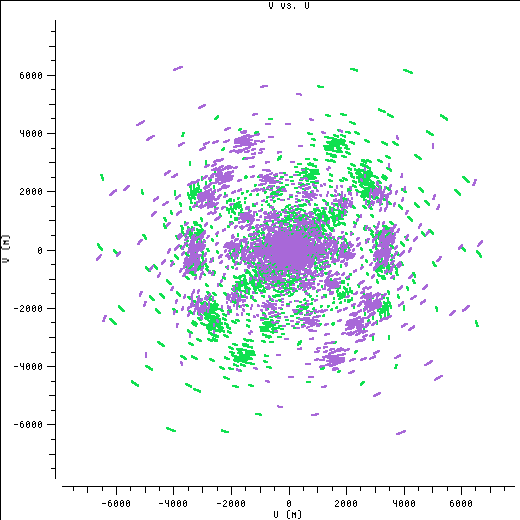}\\
    \includegraphics[width=0.98\columnwidth,clip=True,trim=1 0 0 10]{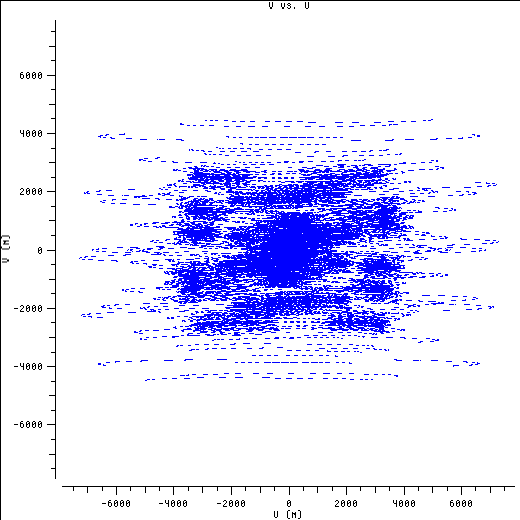}
    \caption{Target \emph{uv}-coverage of the final calibrated data. The effect of MeerKAT's dense core is evident in the well-sampled 0--500\,m \emph{uv}-range, providing maximal sensitivity to extended structure. \textbf{Top panel:} The 24 minute observation of ACT-CL\,J0159.0$-$3413, with the two well-separated 12-minute scans indicated by different colours. \textbf{Bottom panel:} Our longest observation (132 minutes), targeting ACT-CL\,J0240.0$+$0115. The ``flattening'' of the \emph{uv}-tracks is due to the equatorial declination of the target.} 
    \label{fig:uv}
\end{figure}

The observing strategy used in this project breaks up a target observation into 12-minute scans and spreads these over a range of hourangles in order to improve the target \emph{uv}-coverage. Figure \ref{fig:uv} presents the \emph{uv}-coverage plots for one of our shortest observations (ACT-CL\,J0159.0$-$3413 - 24 minutes; top panel) and for our longest observation (ACT-CL\,J0240.0+0115 - 132 minutes; bottom panel). The plotted data are from the final calibrated datasets after all flagging. The effect of spreading out shorter scans produces a wider coverage in the radial direction, as can be seen in the top panel where the two 12-minute scans are in different colours. The effect of MeerKAT's dense core is evident in the well-sampled 0--500\,m range, where there is maximal sensitivity to the extended emission we are searching for. 

\section{Cluster fields with no detections}\label{app:pbimgs}

\begin{figure*}
    \includegraphics[width=0.9\columnwidth,clip=True,trim=0 0 0 0]{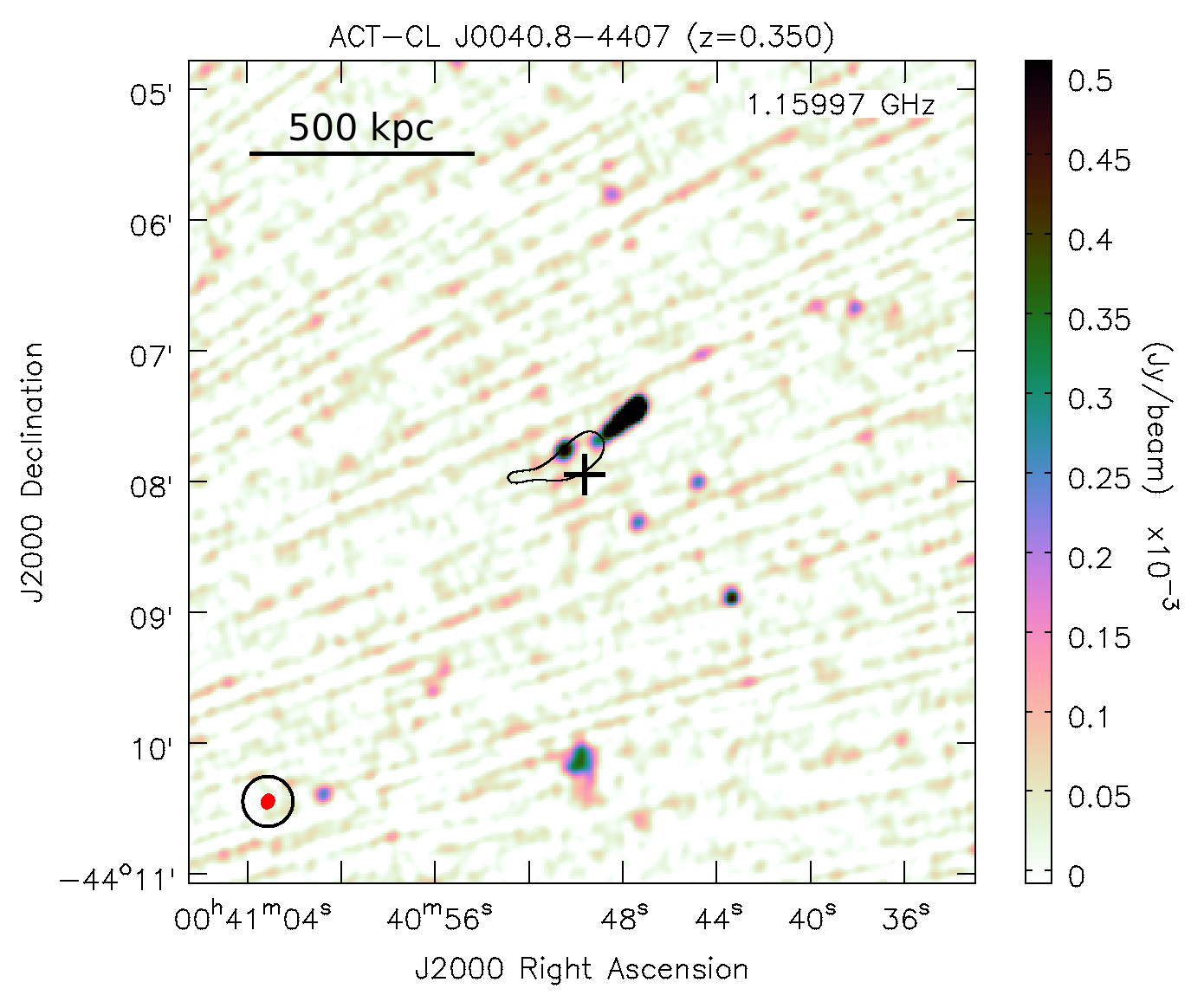}\hspace{1cm} 
    \includegraphics[width=0.9\columnwidth,clip=True,trim=0 0 0 0]{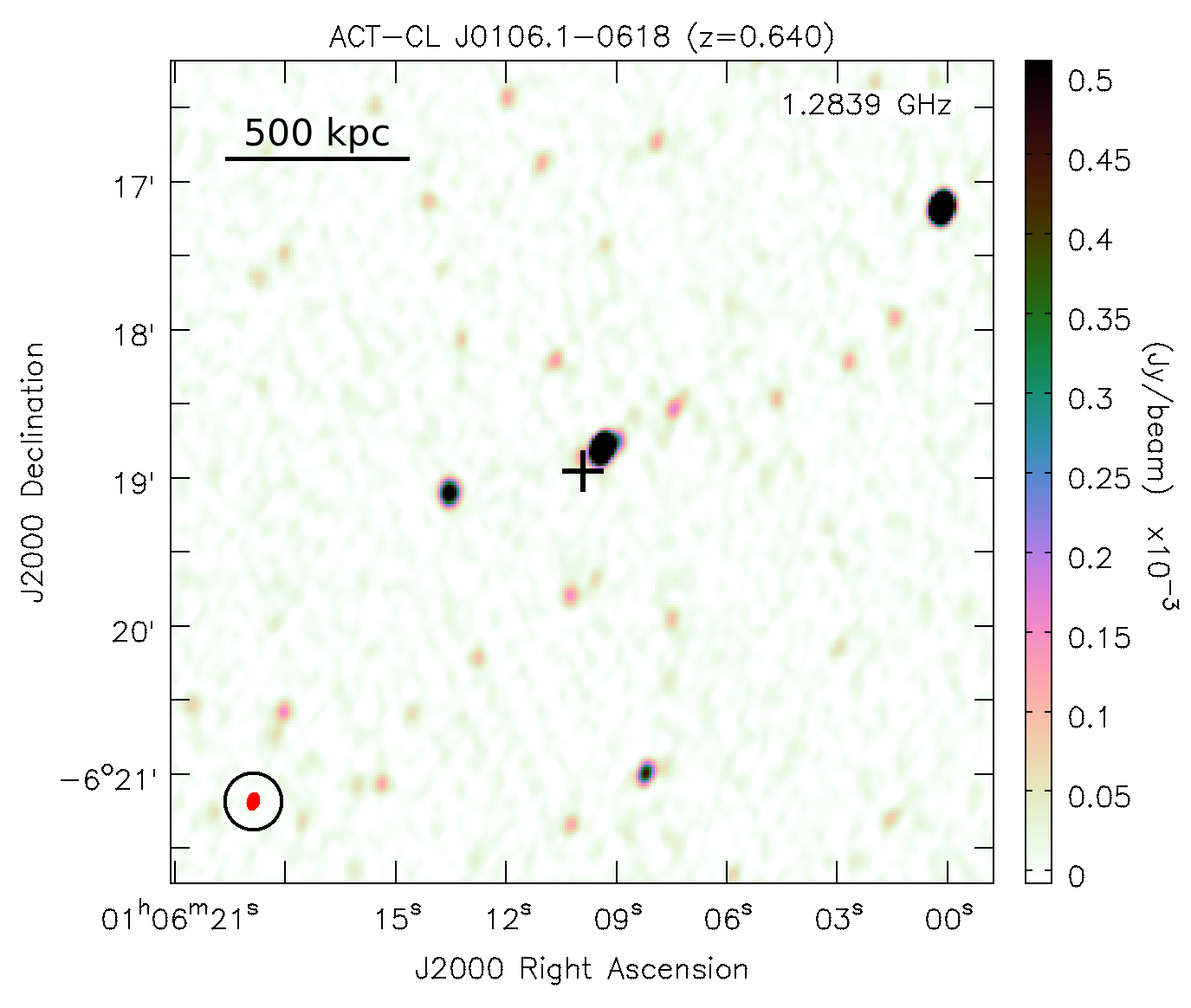}\vspace{0.5cm}
    \includegraphics[width=0.9\columnwidth,clip=True,trim=0 0 0 0]{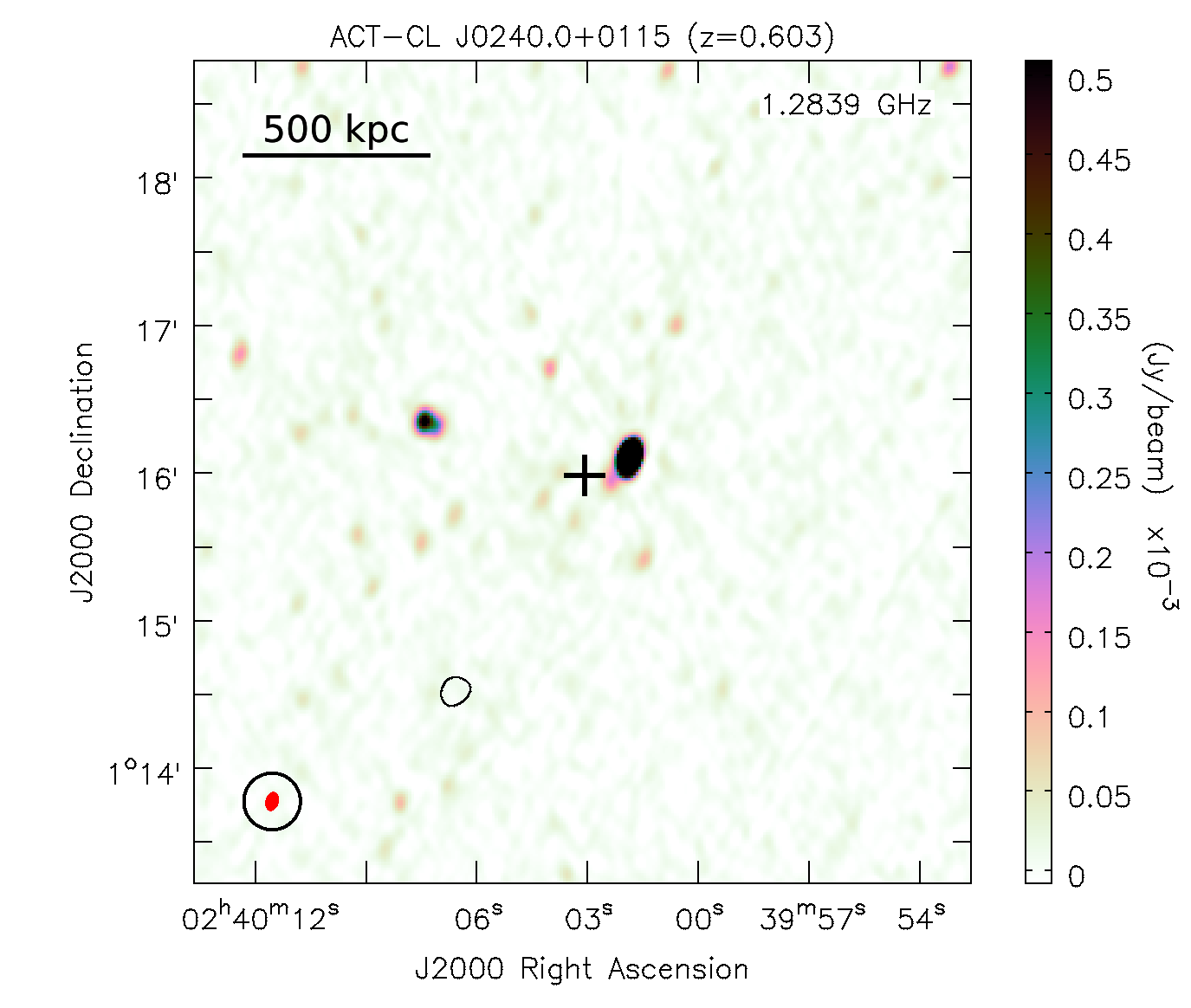}\hspace{1cm}    
    \includegraphics[width=0.9\columnwidth,clip=True,trim=0 0 0 0]{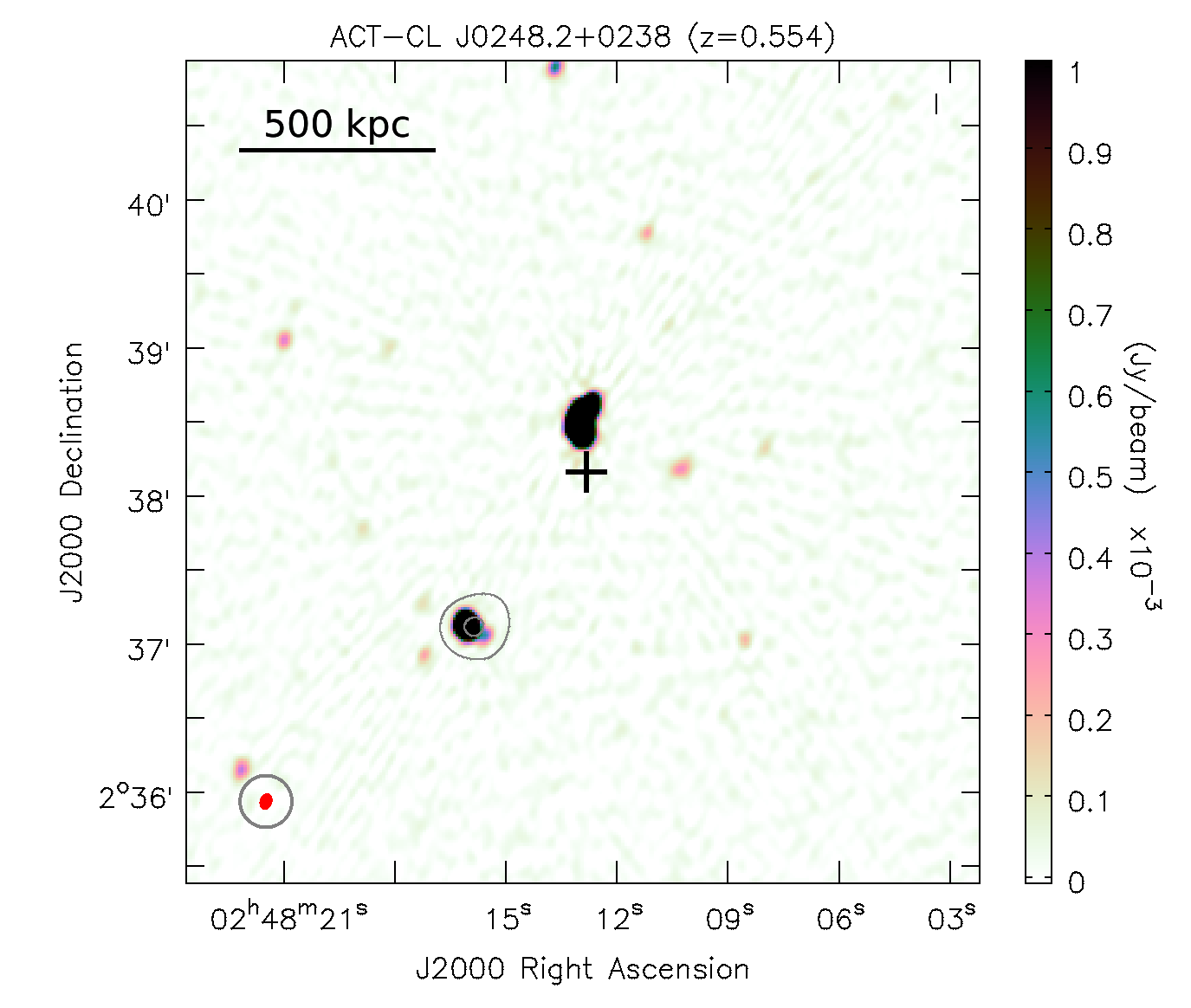}
    \caption{Full-resolution MeerKAT L-band images of the cluster region for the four targets in our sample with no diffuse emission detection. Contours are the 3$\sigma_{\rm LS}$ level from the relevant LS map. The synthesised beam for both the full-resolution (filled red ellipse) and LS maps (open black circle) are indicated in the lower left of each panel. The beam sizes and central rms noise are provided in Table \ref{tab:sampleobs}. The physical scale at the cluster redshift is indicated in the upper left of each panel, and the cross indicates the position of the ACT SZ peak. } 
    \label{fig:nondets}
\end{figure*}

In Figure \ref{fig:nondets} we present full-resolution images of the four targets in our sample with no diffuse emission detection in the cluster region. Colour scales and notations are as in Figure \ref{fig:gallery}. The targets are ACT-CL\,J0040.8$-$4407 (top left), ACT-CL\,J0106.1$-$0618 (top right), ACT-CL\,J0240.0$+$0115 (bottom left), and ACT-CL\,J0248.20$+$0238 (bottom right).


\bsp	
\label{lastpage}
\end{document}